\documentclass[letterpaper,twocolumn,showpacs,preprintnumbers,amsmath,amssymb,nofootinbib,showkeys,aps]{revtex4}
\usepackage{epsfig}

\usepackage{graphicx}
\usepackage{dcolumn}
\usepackage{bm}
\usepackage{amsmath}
\usepackage{latexsym}
\usepackage{color}
\usepackage{hyperref}
\usepackage{mathrsfs}
\newcommand{\be}{\begin{equation}}
\newcommand{\ee}{\end{equation}}
\newcommand{\bea}{\begin{eqnarray}}
\newcommand{\eea}{\end{eqnarray}}

\begin{document}


\title{Exact Solution for Flat Scale-Invariant Cosmology}

\author{J. F. Jesus$^{1,2}$}\email{jfjesus@itapeva.unesp.br}

\affiliation{$^1$Universidade Estadual Paulista (UNESP), Campus Experimental de Itapeva\\R. Geraldo Alckmin, 519, 18409-010, Itapeva, SP, Brazil,
\\$^2$Universidade Estadual Paulista (UNESP), Faculdade de Engenharia, Guaratinguet\'a, Departamento de F\'isica e Qu\'imica - Av. Dr. Ariberto Pereira da Cunha 333, 12516-410, Guaratinguet\'a, SP, Brazil}


\def\zt{\mbox{$z_t$}}

\begin{abstract}
An exact solution for the spatially flat scale-invariant Cosmology, recently proposed by Maeder \cite{Maeder17a} is deduced. No deviation from the numerical solution was detected. The exact solution yields transparency for the dynamical equations and faster cosmological constraints may be performed.

\end{abstract}

\maketitle


\section{Introduction} 
Recently, Maeder \cite{Maeder17a} has proposed that the accelerating expansion of the Universe \cite{SN1,SN2} may be explained by the scale invariance of the empty space. It is argued that in the empty space there is no preferred scale of length or time. He applies the theoretical framework developed for a scale-invariant theory of gravitation \cite{Weyl23,Eddington23,Dirac73,CanutoEtAl77} and finds, as a result, that the effects of scale invariance are smaller when there is greater quantity of matter present in the Universe.

This model has some interesting consequences. The cosmological constant, for example, is replaced by a variable term involving the scale-invariant scale factor. There is no need for cosmological constant nor dark energy in this context. The scale-invariant effects over the CMB temperature $T_{CMB}(z)$ were analysed in \cite{Maeder17b}, where he finds that if Galactic corrections are applied, scale invariance may not be discarded. By analysing the Milky Way rotation curve in the context of scale invariant model, \cite{Maeder17c} argues that the flat rotation curves may be seen as an age effect and indicates that nor dark matter is needed in this theoretical context.

Here, it is focused in finding exact solutions of his equation of cosmological dynamics. Exact solutions provides transparency for the cosmological equations, may provide insights and can be used for faster cosmological constraints.

In section \ref{scaleinv} the equations of scale-invariant Cosmology are briefly discussed. In section \ref{scaleflat}, the exact solution for spatially flat scale-invariant Cosmology is deduced and compared with the numerical results by \cite{Maeder17a}. In section \ref{conclusion}, the conclusions are presented.

\section{\label{scaleinv}Scale-Invariant Cosmology} 
Maeder \cite{Maeder17a} shows that the general relativity (GR) metric is related to the scale-invariant one by $ds'=\lambda(x^\mu)ds$, where $\lambda$ is the scale factor which connects both line elements. By applying this theory to the empty space, he finds that $\lambda$ relates to the Einstein's cosmological constant, $\Lambda_E$, by
\be
\lambda=\sqrt{\frac{3}{\Lambda_E}}\frac{1}{ct}
\ee
where $c$ is the speed of light, which will be assumed to unity elsewhere.

By applying the Robertson-Walker metric to a Universe filled with pressureless matter, he finds for the scale-invariant Cosmology:
\be
\dot{R}^2Rt - 2\dot{R}R^2 + kRt - Ct^2=0
\label{odeRk}
\ee
where $R$ is the scale factor, times is expressed in units of $t_0=1$, $k$ is the curvature constant, which can be $0$, $-1$ or $+1$ for spatially flat, open or closed Universe, respectively, $C=\frac{8\pi G\rho_mR^3\lambda}{3}$ and $\rho_m$ is density of pressureless matter. The density parameters are defined in the usual form, $\Omega_m\equiv\frac{\rho}{\rho_c}$, where $\rho_c=\frac{3H^2}{8\pi G}$, $\Omega_k=-\frac{k}{R^2H^2}$, but $\Omega_\Lambda$ is now replaced by $\Omega_\lambda$:
\be
\Omega_\lambda\equiv\frac{2}{Ht}
\label{wl}
\ee
such that it is valid the normalization condition:
\be
\Omega_m+\Omega_k+\Omega_\lambda=1.
\label{norm}
\ee

Our focus here is to find exact solutions of (\ref{odeRk}). As nonflat exact solutions could not be found, we aim to solve the case where $k=0$.

\section{\label{scaleflat}Flat Scale-Invariant Cosmology}
For the spatially flat Universe, the equation for scale-invariant Universe evolution (\ref{odeRk}) reads:
\be
\dot{R}^2Rt - 2\dot{R}R^2-Ct^2=0
\label{odeR}
\ee
where time is expressed in units of $t_0=1$, and it is assumed that $R_0=1$. By evaluating (\ref{odeR}) today, the relation for the Hubble constant $H_0=\frac{\dot{R}_0}{R_0}$ is $H_0^2-2H_0=C$. It yields $H_0=1\pm\sqrt{1+C}$ and we choose the positive sign since the Universe is expanding ($H_0>0$).

In order to solve (\ref{odeR}), we replace $R(t)$ by $R(t)=tv(t)$, so:
\be
\dot{R}=v+t\dot{v}
\label{dotRv}
\ee
and (\ref{odeR}) may be written as:
\be
\dot{v}^2=\frac{v^3+c}{vt^2}\Rightarrow\dot{v}=\pm\sqrt{\frac{v^3+c}{vt^2}}
\label{odev}
\ee
So, we have to choose a sign for $\dot{v}$ in order to integrate the separable equation (\ref{odev}). From (\ref{dotRv}), we have:
\be
\dot{v}=\frac{R}{t^2}(Ht-1)
\ee

In such a way that we have $\dot{v}_0=H_0-1=\sqrt{1+C}$. So, we choose the positive sign in (\ref{odev}) in order to integrate it from today. The result is:
\be
\sqrt{v^3+C}+v^{3/2}=c_1t^{3/2}
\label{eqvt}
\ee
where $c_1$ is an integration constant. From (\ref{eqvt}), we find
\be
c_1=1+\sqrt{1+C}=H_0
\label{c1}
\ee

We want to solve (\ref{eqvt}) for $v(t)$ in order to write $R(t)$. We write $v^{3/2}=c_1t^{3/2}-\sqrt{v^3+C}$ and square both sides. The result is:
\be
v=\left[\left(\frac{c_1^2t^3-C}{2c_1t^{3/2}}\right)^2\right]^{1/3}
\label{v2}
\ee

Before writing the power as $2/3$ in (\ref{v2}), there was an ambiguity sign inside the parentheses after we squared the expression (\ref{eqvt}). We must choose the sign which is in agreement with the initial condition (\ref{c1}). By choosing the negative sign inside the parentheses, we find $c_1=-1\pm\sqrt{1+C}$, which are both in disagreement with (\ref{c1}). By choosing the positive sign, we find $c_1=1\pm\sqrt{1+C}$, where there is a solution in agreement with (\ref{c1}), so we choose the positive sign and:
\be
v=\left(\frac{c_1^2t^3-C}{2c_1t^{3/2}}\right)^{2/3}
\ee
Now:
\be
R=tv=\left[\frac{(2H_0+C)t^3-C}{2H_0}\right]^{2/3}
\label{Rct}
\ee
where we have replaced the value of $c_1$. As explained by \cite{Maeder17a} and shown in Eqs. (\ref{wl})-(\ref{norm}), in flat scale invariant Cosmology, $\Omega_m$ can be written today as:
\be
\Omega_m+\frac{2}{H_0t_0}=1
\ee
so $H_0=\frac{2}{1-\Omega_m}$ in units of $t_0$ and $C=\frac{4\Omega_m}{(1-\Omega_m)^2}$. So, by writing (\ref{Rct}) in terms of $\Omega_m$:
\be
R=\left[\frac{t^3-\Omega_m}{1-\Omega_m}\right]^{2/3}
\label{Rt}
\ee

\begin{figure}[ht]
 \includegraphics[width=\linewidth]{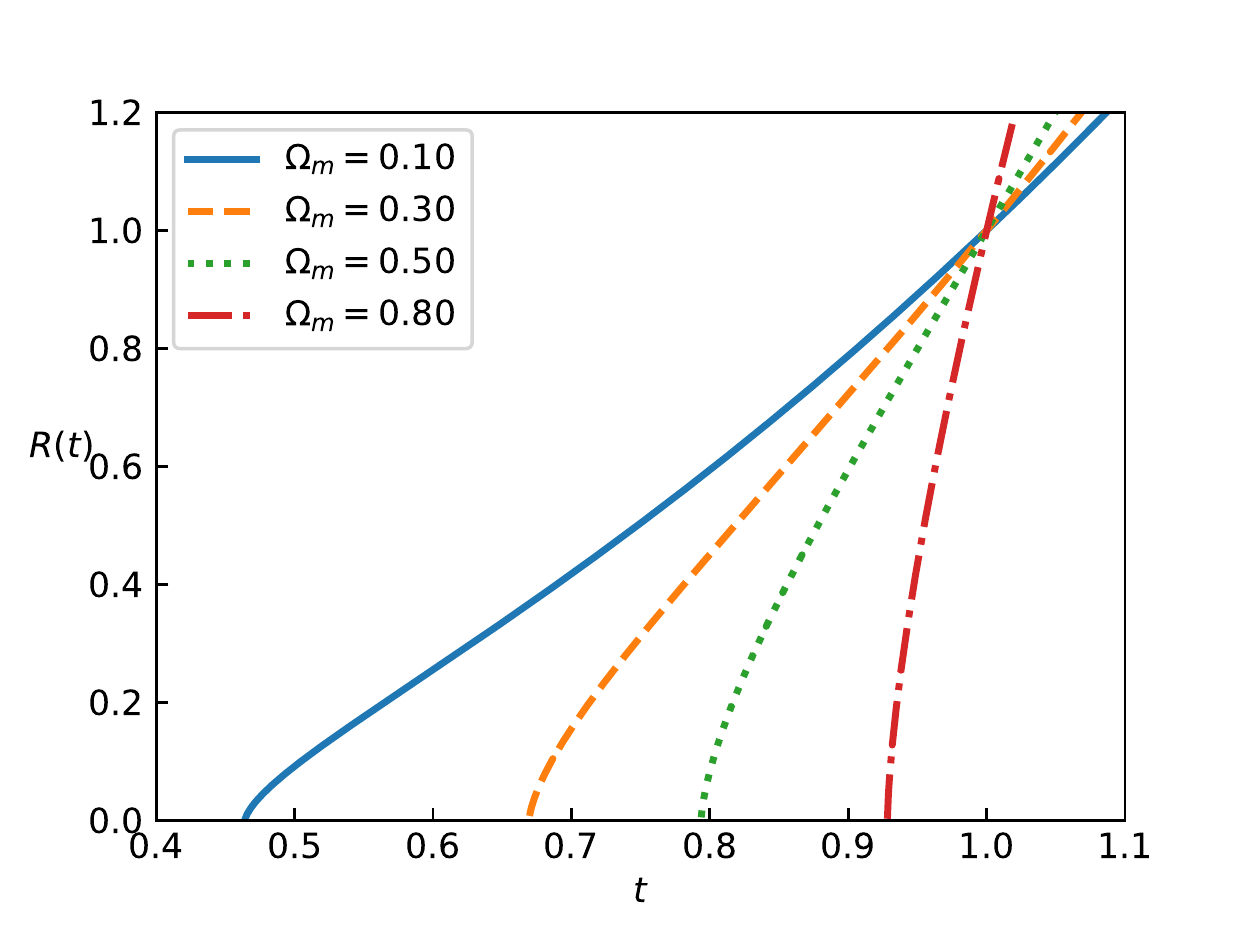}
 \caption{\label{Rtwm} Scale factor $R(t)$ for some values of the matter density parameter in the flat scale-invariant Cosmology.}
\end{figure}

In Fig. \ref{Rtwm}, we show some evolutions of the scale factor for different values of the density parameter, based in Eq. (\ref{Rt}). It may be compared with Fig. 2 of \cite{Maeder17a}, where the solutions were numerically obtained. No difference may be found between the exact and numerical results.

As can be seen in Fig. \ref{Rtwm} and Eq. (\ref{Rt}), with time in units of $t_0$ there is an initial time, $t_{in}$, where $R=0$. From (\ref{Rt}), this time is:
\be
t_{in}=t_0\Omega_m^{1/3}.
\ee
From this we may write for the total age of the Universe, $\tau_0=t_0-t_{in}$:
\be
\tau_0=(1-\Omega_m^{1/3})t_0=\frac{2}{H_0}\frac{1-\Omega_m^{1/3}}{1-\Omega_m}
\ee

By solving (\ref{Rt}) for time, we may write for the age at redshift $z$, $\tau(z)=t(z)-t_{in}$:
\be
\tau(z)=\frac{2}{H_0(1-\Omega_m)}\{[\Omega_m+(1-\Omega_m)(1+z)^{-3/2}]-\Omega_m^{1/3}\}
\label{tauz}
\ee

By calculating $H=\frac{\dot{R}}{R}$ from (\ref{Rt}), we find:
\be
H(t)=\frac{2t^2}{t^3-\Omega_m}
\label{Ht}
\ee

We solve Eq. (\ref{Rt}) for time, then replace on (\ref{Ht}) to find:
\bea
H&=&\frac{2[\Omega_m+(1-\Omega_m)R^{3/2}]^{2/3}}{(1-\Omega_m)R^{3/2}}\nonumber\\
&=&H_0[\Omega_mR^{-9/4}+(1-\Omega_m)R^{-3/4}]^{2/3}
\eea
or, in terms of redshift $z$:
\be
H=H_0[\Omega_m(1+z)^{9/4}+(1-\Omega_m)(1+z)^{3/4}]^{2/3}
\label{Hz}
\ee

\begin{figure}[ht]
 \includegraphics[width=\linewidth]{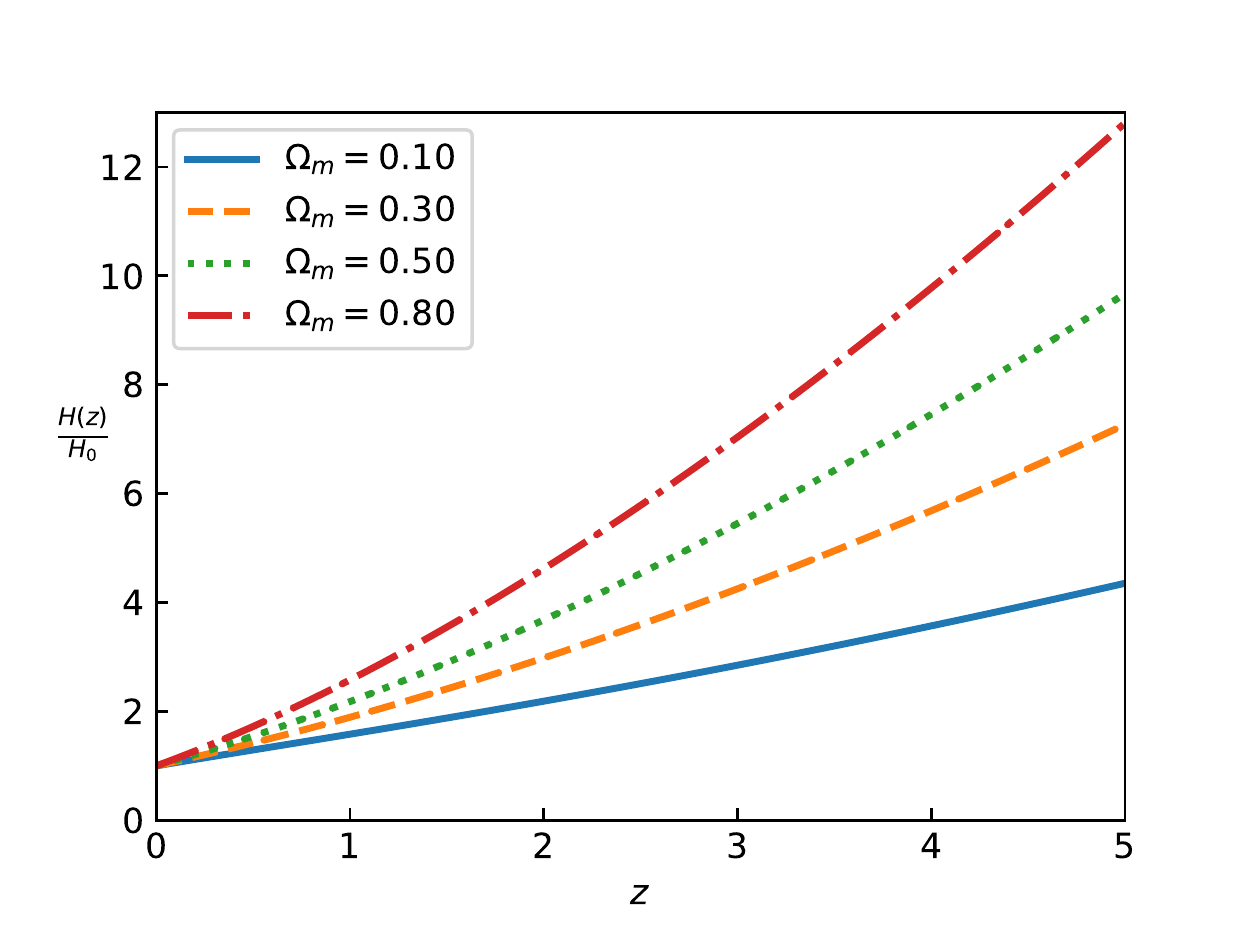}
 \caption{\label{Hzwm} Hubble parameter $H(z)$ for some values of the matter density parameter in the flat scale-invariant Cosmology.}
\end{figure}

In Fig. \ref{Hzwm}, it is plotted $H(z)$ for some values of the matter density parameter from Eq. (\ref{Hz}).

From Eqs. (\ref{Rt}), (\ref{tauz}) and (\ref{Hz}) the results of Tables 1 and 2 of Ref. \cite{Maeder17a} can be recovered. No difference between analytical and numerical results could be found.

\newpage
\section{\label{conclusion}Conclusion}

An interesting theory of scale-invariant Cosmology was recently proposed. While the original focus were in the numerics, here we focus in finding an exact solution, at least for the spatially flat case. We show that our exact solution does not deviate from the original numerical solution. Physical insights about the dynamical equations may now be developed and faster cosmological constraints can be made.

\begin{acknowledgments}
JFJ is supported by Funda\c{c}\~ao de Amparo \`a Pesquisa do Estado de S\~ao Paulo - FAPESP (Processes no. 2013/26258-4 and 2017/05859-0).
\end{acknowledgments}


\end{document}